\documentclass{PoS}
\usepackage{mathrsfs}
\usepackage{subfigure}
\usepackage{verbatim}

\def\mpt{{\slash\!\!\!\!\!\:P}_T}

\def\mptv{{\slash\!\!\!\!\!\:\vec{P}}_T}

\title{Phenomenology of the minimal B-L extension of the Standard Model}

\ShortTitle{Phenomenology of the minimal B-L extension of the SM}

\author{\speaker{L.~Basso}$^{,1}$, A.~Belyaev$^1$, S.~Moretti$^{1,2}$, G.M.~Pruna$^1$ and C.H.~Shepherd-Themistocleous$^{1,}$\thanks{This work is supported in part by the SEPnet Physics Consortium. SM is financially supported in part by 
the scheme `Visiting Professor - Azione D - Atto Integrativo tra la 
Regione Piemonte e gli Atenei Piemontesi.}\\
        $^1$NExT Institute, University of Southampton, Highfield, Southampton, 
        and RAL, Chilton, Didcot, ~UK\\
        $^2$Dipartimento di Fisica Teorica, Universit\`a di Torino, Torino, Italy\\
        E-mail: \email{lb4x07@soton.ac.uk}, ~\email{a.belyaev@soton.ac.uk}, \email{s.moretti@soton.ac.uk},~\email{g.m.pruna@soton.ac.uk},~\email{claires@mail.cern.ch}}

\abstract{We present the Large Hadron Collider (LHC) discovery
potential in the $Z'$ and heavy neutrino sectors of a $U(1)_{B-L}$
enlarged Standard Model also encompassing three heavy Majorana
neutrinos. This model exhibits novel signatures at the LHC,
the most interesting arising from a $Z'$ decay chain involving
heavy neutrinos, eventually decaying into leptons and jets. In
particular, this signature allows one to measure the $Z'$ and heavy
neutrino masses involved.  In addition, over a large region of
parameter space, the heavy neutrinos are rather long-lived particles
producing distinctive displaced vertices that can be 
seen in the detectors.  Lastly, the simultaneous
measurement of both the heavy neutrino mass and decay length enables
an estimate of the absolute mass of the parent light neutrino.
For completeness, we will also compare the LHC and a future 
Linear Collider (LC) discovery potentials.}

\FullConference{European Physical Society Europhysics Conference on High Energy Physics\\
                 July 16-22, 2009\\
                 Krakow, Poland}

\begin{document}

\section{Introduction}
The 
$B-L$ (baryon number minus lepton number) symmetry plays an important role in
various physics scenarios beyond the Standard Model (SM). First, 
the gauged  $U(1)_{B-L}$ symmetry group is 
contained in {several} Grand Unified Theories.
%, {such as the one based on the} $SO(10)$
% group~\cite{Buchmuller:1991ce} (which can have an intermediate symmetry
% ${SU(2)}_L \times {SU(2)}_R \times {SU(3)}_C \times {U(1)}_{B-L}$ 
%which breaks down to the group of the SM at a scale of $10^{10-12}$ GeV).
Second, the scale of  
the $B-L$ symmetry  breaking is related to the mass scale of the heavy right-handed 
Majorana neutrino mass terms providing the well-known see-saw
mechanism of light neutrino mass generation. Third, the
$B-L$ symmetry and the scale of its breaking are tightly connected to
the baryogenesis mechanism through leptogenesis
via sphaleron interactions preserving $B-L$.

In this work, we study in detail the collider phenomenology of the gauge and fermionic sectors of the minimal $B-L$ extension of the SM, i.e., the $Z'$ and the 
three heavy neutrinos and their interplay, in the framework of the LHC. A very interesting feature of such a $B-L$ model is possibly relatively long lifetimes of the heavy neutrinos, which can directly be measured. Such measurement could also be a key to shedding light on the mass spectra of the light neutrinos.

\section{The model}
The model under study is the so-called ``pure'' or ``minimal''
$B-L$ model (see \cite{bbms} for conventions and references) 
since it has vanishing mixing between the two $U(1)_{Y}$ 
and $U(1)_{B-L}$ groups.
%In the rest of this work we refer to this model simply as the ``$B-L$ model''. 
In this model the classical gauge invariant Lagrangian,
obeying the $SU(3)_C\times SU(2)_L\times U(1)_Y\times U(1)_{B-L}$
gauge symmetry, can be decomposed as usual as $\displaystyle
\mathscr{L}=\mathscr{L}_{YM} + \mathscr{L}_s + \mathscr{L}_f + \mathscr{L}_Y$.
The non-Abelian field strengths in $\mathscr{L}_{YM}$ are the same as in the SM
whereas the Abelian ones can easily be diagonalised\footnote{In general, Abelian field strengths tend to mix and the diagonalisation of the kinetic terms could be complicated. However, in our case just one off-diagonal term appears and a linear $2\times 2$ transformation is sufficient to fulfill our aim.}.
In such field basis, the covariant derivative is:
%\begin{equation}\label{cov_der}
$\displaystyle D_{\mu}\equiv \partial _{\mu} + ig_S T^{\alpha}G_{\mu}^{\phantom{o}\alpha} 
+ igT^aW_{\mu}^{\phantom{o}a} +ig_1YB_{\mu} +i(\widetilde{g}Y + g_1'Y_{B-L})B'_{\mu}\, .$
%\end{equation}
The ``pure'' or ``minimal'' $B-L$ model is defined by the condition $\widetilde{g} = 0$, that implies no mixing between the $Z'$ and the SM $Z$ gauge bosons.

The fermionic Lagrangian is the same as in the SM, apart from the term associated to RH-neutrinos ($i\overline {\nu _{kR}} \gamma _{\mu}D^{\mu} \nu_{kR}$, where $k$ is the
generation index). The fields' charges are the usual SM and $B-L$ ones (in particular, $B-L = 1/3$ for quarks and $-1$ for leptons).
  The latter charge assignments as well as the introduction of three new
  fermionic  right-handed heavy neutrinos ($\nu_R$) and one
  scalar Higgs ($\chi$, charged $+2$ under $B-L$)  
  fields are designed to eliminate the triangular $B-L$  gauge anomalies and to ensure the gauge invariance of the theory, respectively.
  Therefore, the $B-L$  gauge extension of the SM group
  broken at the Electro-Weak (EW) scale does necessarily require
  at least one new scalar field and three new fermionic fields which are
  charged with respect to the $B-L$ group.

Regarding the scalar Lagrangian, the only differences with respect to the SM is that we have to introduce a kinetic term for the $\chi$ field as well as to modify the scalar potential, given by
%\begin{equation}\label{new-potential}
$\displaystyle V(H,\chi ) = m^2H^{\dagger}H +
 \mu ^2\mid\chi\mid ^2 +
  \lambda _1 (H^{\dagger}H)^2 +\lambda _2 \mid\chi\mid ^4 + \lambda _3 H^{\dagger}H\mid\chi\mid ^2  \, ,$
%\end{equation}
{where $H$ and $\chi$ are the complex scalar Higgs 
doublet and singlet fields, respectively.}

Finally, as for Yukawa interactions, we are allowed to introduce two new terms ($\displaystyle y^{\nu}_{jk}\overline {l_{jL}} \nu _{kR}$ $\widetilde H 
	         + y^M_{jk}\overline {(\nu _R)^c_j} \nu _{kR}\chi$, where $\widetilde H=i\sigma^2 H^*$ and  $i,j,k$ take the values $1$ to $3$),
where the first term is the usual Dirac contribution and the second term is the Majorana one.
Neutrino mass eigenstates, obtained after applying the see-saw mechanism, will be called $\nu_l$ and $\nu_h$, being the former the SM-like ones. With a reasonable choice of the Yukawa couplings, the heavy neutrinos $\nu_h$ can have masses $m_{\nu_h} \sim \mathcal{O}(100)$ GeV, within the LHC reach. Their role will be discussed later on.

\section{Discovery potential at the LHC and at future LC}
We want to explore the discovery potential of hadronic and leptonic machines in the $M_{Z'}$-$g_1'$ plane of our model, in the di-muon production process.
We compare the LHC hadronic scenario, with $100$ fb$^{-1}$ data collected, to two different LC leptonic frameworks, at a fixed Centre-of-Mass
(CM) energy of $\sqrt{s_{e^+e^-}}=3$ TeV ($500$ fb$^{-1}$ data altogether) and in a so-called energy scan, where the CM energy is set to $\sqrt{s_{e^+e^-}}=M_{Z'}+10$ GeV and we assume $10\mbox{ fb}^{-1}$ of luminosity for each step. 
%We compare $100$ fb$^{-1}$ data collected at the LHC with $500$ fb$^{-1}$ at a LC with a fixed CM energy of $\sqrt{s_{e^+e^-}}=3$ TeV and $10$ fb$^{-1}$ per step in a so-called energy scan approach at a LC.
%and again a LC in a so-called energy scan approach, where the CM energy  is set to $\sqrt{s_{e^+e^-}}=M_{Z'}+10$ GeV (assuming $10\mbox{ fb}^{-1}$ of luminosity for each step).
We then limit both signal and background to the
detector acceptance volumes and $M_{\mu\mu}$ to an invariant mass window defined by the CMS and ILC prototype resolution \cite{CMS-ILCdet} or $3\Gamma_{Z'}$, whichever the largest. We finally define the \emph{significance} $\sigma$ as $s/ \sqrt{b}$ ($s$ and $b$ being the signal and background event rates, respectively)\footnote{This definition, based on a gaussian distribution, is valid when the number of events is large enough, i.e. $s,\,b>20$. Otherwise, in case of lower statistics, we exploited the Bityukov algorithm \cite{Bityukov}, which uses the Poisson `true' distribution.}: the discovery will be for $\sigma\ge 5$, as usual.

As a result,
for $M_{Z'}>800$ GeV, the LC potential to explore the $M_{Z'}$-$g_1'$
parameter space in the fixed CM energy approach goes beyond the LHC reach.
For example, for $M_{Z'}=1$ TeV, the
LHC can discover a $Z'$ if $g'\approx 0.007$
while a LC can achieve this for $g'\approx 0.005$.
The difference is even  more drastic for larger $Z'$ masses:
 a LC can discover a $Z'$ with a $2$ TeV mass for a 
$g_1'$ coupling which is a factor 8 smaller.
% than the one for which the same mass $Z'$ can be discovered at the LHC.

In case of the energy scan approach,
the $M_{Z'}$-$g_1'$ parameter space can be probed even further
for $M_{Z'}<1.75$~TeV.
For example, for $M_{Z'}=1$~TeV, 
$g_1'$ couplings can be probed down to the $2.6\times 10^{-3}$,
following a $Z'$ discovery.
Furthermore, the parameter space 
corresponding to the mass interval  $500~{\rm GeV}~<M_{Z'}<1$ TeV, 
which the LHC covers better as compared to 
 a LC with fixed energy, can be accessed well beyond the LHC reach 
with a LC in energy scan regime.
%Altogether then, both an ILC, $\sqrt{s_{e^+e^-}}\leq 1$ TeV) \cite{ILC} and a 
%Compact Linear Collider (CLIC, $\sqrt{s_{e^+e^-}}\leq 3$ TeV) \cite{CLIC}
%design may be able (over suitable regions of $B-L$ parameter space) to outperform the LHC.

\section{$Z'$ and neutrino phenomenology}
The possibility of the $Z'$ gauge boson decaying into pairs of heavy
neutrinos is one of the most significant results of this work
since, in addition to the clean SM-like di-lepton signature, it
provides multi-lepton signatures where backgrounds can strongly be 
suppressed.

In order to address this quantitatively, we first
determine the relevant Branching Ratios (BRs): clearly, these depend strongly
on the heavy neutrino mass.
A feature of the current $B-L$ model is that the $Z'$ predominantly couples to
leptons. In fact, after summing over the generations, we
roughly get for leptons a total BR of $3/4$ and for quarks the remaining $1/4$.
Not surprisingly then, for a relatively light (with respect to the $Z'$ gauge boson) heavy neutrino, the $Z'$ BR into pairs of such particles is relatively high: $\sim 20\% $ (at most, after summing over the generations).
Regarding finally the total $Z'$ width, it strongly depends on the coupling up to few hundreds GeV for a TeV scale gauge boson.
%%%%%%%%%%%%%%%%%%%%%%%%%%%%%%%%%%%%%%%%%%%%%%

Moving to the neutrino sector, after the see-saw diagonalisation of the neutrino mass matrix, we obtain three very light neutrinos ($\nu_l$), which are the SM-like neutrinos, and three heavy neutrinos ($\nu_h$).
The latter have an extremely small mixing
with the  $\nu_l$'s thereby providing very small but non-vanishing 
couplings to gauge and Higgs bosons. 
Hence, neglecting the scalar sector, the $\nu_h$'s prefer to decay into SM gauge bosons, as well as into the new $Z'$, when these decay channels are kinematically allowed.
% which in turn enable the $\nu_h$ to dominantly decay into SM gauge bosons (neglecting the scalar sector), as well as into the new $Z'$ when these decay channels are kinematically allowed.
In details, BR$\left( \nu_h \rightarrow l^\mp W^\pm \right)$
is dominant and reaches the $2/3$ level in the  $M_{Z'}> M_{\nu_h} \gg M_W, M_Z$
limit, while the BR$\left( \nu_h\rightarrow \nu_l Z \right)$ represents the remaining $1/3$ in this regime. 
%In contrast, the $\nu_h \rightarrow \nu_l h_2$ as well as $\nu_h \rightarrow \nu_l Z'$ decay channels are well below the percent level and are negligible.

The $\nu_h$ couplings to the gauge bosons are
proportional to the ratio of light to heavy neutrino masses, which is extremely small.  Therefore the decay width of
the heavy neutrino is correspondingly small and its lifetime large: it can be a long lived particle and, over a
large portion of parameter space, its lifetime can be comparable to or
exceed that of the $b$-quark, giving rise to a displaced vertex inside the detector.
The key point is that a measurable lifetime along with a mass
determination for $\nu _h$ also enables a determination of
$m_{\nu l}$, by just inverting the see-saw formula.
%The lifetime measurement allows the small
%heavy-light neutrino mixing angle to be determined and 
%this, along with $m_{\nu h}$, gives the light neutrino mass, by mean of: $\displaystyle \tan{2 \alpha _\nu} = -2 \sqrt{\frac{m_{\nu l}}{m_{\nu h}}}$.

\section{Heavy neutrino mass measure at the LHC}
Multi-lepton signatures carry the hallmark of the heavy neutrinos
as the latter enter directly the corresponding decay chains.
%and the further leptons come from the $W$ and $Z$ following decays.
We performed a detailed Monte Carlo analysis
at the benchmark point $M_Z' = 1.5$ TeV, $g'_1 = 0.2$
and $M_{\nu_h}=200$ GeV, with $\sigma (pp\rightarrow \nu _h \nu_h) = 46.7$ fb (for CTEQ6L PDFs with $Q^2=M_{Z'}^2$), and the decay we are interested in is $\nu _h \nu_h \rightarrow 3l+2q+\nu _l$, with a significant fraction of missing energy. A very suitable distribution to look at turned out to be the transverse mass defined in \cite{Barger}, i.e.,
%\begin{equation}
$\displaystyle m^2_T = \left( \sqrt{M^2(vis)+P^2_T(vis)}+\left| \mpt \right| \right) ^2
	- \left( \vec{P}_T(vis) + \mptv\right) ^2\, $,
%\end{equation}
where $(vis)$ means {the sum over} the visible particles.  If the
visible particles we sum over are the $3$ leptons and $2$
jets, the transverse mass distribution will peak at the $Z'$ mass. We can
also see evidence for the presence of a heavy neutrino by just
considering as visible particles the $2$ leptons with the smallest 
azimuth-rapidity separation, since this is the topology relevant to a $\nu_h$ decay.  The
results show that this transverse mass peak for the heavy neutrino is
likely to be the best way to measure its mass. The striking signature of
this model is that both of the above peaks occur simultaneously.

It is important to note that the backgrounds are completely under control: as sources we considered $WZjj$, $t\overline{t}$ (where a further lepton comes from a semileptonic $b$ decay) and $t\overline{t}l\nu$. After simple kinematic and detector acceptance requirements, significant suppression of the backgrounds comes from requiring the di-jet invariant mass to be close to the $W$ mass (as this is the signal topology). At this stage, just the $WZjj$ is comparable to our signal, but the $Z'$ peak described above can be used to further reduce the remnant background.

\begin{comment}
\section{Conclusions}
We have analyzed the LHC discovery potential in the $Z'$ and heavy
neutrino sector of a (broken) $U(1)_{B-L}$ enlarged SM also
encompassing three heavy (Majorana) neutrinos and found that novel
signals can be established. The most interesting {new signature
involves} three leptons (electron and/or muons), two jets plus missing
transverse momentum {coming from a $Z'$ decay chain into heavy
neutrinos}. Various mass distributions (both invariant and
transverse) can be used to not only extract the signal after a few
years of LHC running, but also to measure the $Z'$ and heavy neutrino
masses involved. This is possible through DY production and decay via
$q\bar q\to Z'\to \nu_h \nu_h$.  In fact, for a large portion of the
parameter space {of our $B-L$ model}, the heavy neutrinos are rather
long-lived particles, so that they produce displaced vertices in the
LHC detectors. 

In addition, from the simultaneous measurement of both the heavy
neutrino mass and decay length one can estimate the {absolute} mass
of the parent light neutrino, for which at present, only limits
exist. 

A study of the Higgs sector is in progress, related to the phenomenology and bounds analysis as well as the possibility of using the heavy neutrino as a source of Higgs via $\nu _h \rightarrow \nu _l h1$.
\end{comment}

\end{document}